\documentclass[apjl]{emulateapj}
\usepackage{apjfonts}

\newcommand{\galex}{{\it GALEX~}}

\slugcomment{Received 2004 May 10; accepted 2004 Jul 29}
\journalinfo{Astrophysical Journal Letters Special ({\it GALEX}) Issue (in press)}

\shorttitle{STAR FORMATION HISTORIES FROM \galex}
\shortauthors{SALIM ET AL.}

\begin{document}

\title{New Constraints on the Star Formation Histories and Dust Attenuation \\
of Galaxies in the Local Universe from  \galex}

\author{
Samir Salim\altaffilmark{1},
St\'ephane Charlot\altaffilmark{2,11},
R.\ Michael Rich\altaffilmark{1},
Guinevere Kauffmann\altaffilmark{2},
Timothy M.\ Heckman\altaffilmark{3},
Tom A.\ Barlow\altaffilmark{4},
Luciana Bianchi\altaffilmark{5},
Yong-Ik Byun\altaffilmark{6}, 
Jose Donas\altaffilmark{7},
Karl Forster\altaffilmark{4},
Peter G.\ Friedman\altaffilmark{4},
Patrick N.\ Jelinsky\altaffilmark{8},
Young-Wook Lee\altaffilmark{6},
Barry F.\ Madore\altaffilmark{9},
Roger F.\ Malina\altaffilmark{7},
D.\ Christopher Martin\altaffilmark{4},
Bruno Milliard\altaffilmark{7},
Patrick Morrissey\altaffilmark{4},
Susan G.\ Neff\altaffilmark{10},
David Schiminovich\altaffilmark{4},
Mark Seibert\altaffilmark{4},
Oswald H.\ W.\ Siegmund\altaffilmark{8},
Todd Small\altaffilmark{4},
Alex S.\ Szalay\altaffilmark{3},
Barry Y.\ Welsh\altaffilmark{8}, and
Ted K.\ Wyder\altaffilmark{4}
}
\email{samir@astro.ucla.edu}

\altaffiltext{1}{Department of Physics and Astronomy, University of
California, Los Angeles, CA 90095}

\altaffiltext{2}{Max-Planck Institut f\"ur Astrophysik,  
D-85748 Garching, Germany}

\altaffiltext{3}{Department of Physics and Astronomy, The Johns Hopkins
University, Homewood Campus, Baltimore, MD 21218}

\altaffiltext{4}{California Institute of Technology, MC 405-47, 1200 East
California Boulevard, Pasadena, CA 91125}

\altaffiltext{5}{Center for Astrophysical Sciences, The Johns Hopkins
University, 3400 N. Charles St., Baltimore, MD 21218}

\altaffiltext{6}{Center for Space Astrophysics, Yonsei University, Seoul
120-749, Korea}

\altaffiltext{7}{Laboratoire d'Astrophysique de Marseille, BP 8, Traverse
du Siphon, 13376 Marseille Cedex 12, France}

\altaffiltext{8}{Space Sciences Laboratory, University of California at
Berkeley, 601 Campbell Hall, Berkeley, CA 94720}

\altaffiltext{9}{Observatories of the Carnegie Institution of Washington,
813 Santa Barbara St., Pasadena, CA 91101}

\altaffiltext{10}{Laboratory for Astronomy and Solar Physics, NASA Goddard
Space Flight Center, Greenbelt, MD 20771}

\altaffiltext{11}{Institut d'Astrophysique de Paris, CNRS,
98 bis boulevard Arago, F-75014 Paris, France}

\begin{abstract}

We derive a variety of physical parameters including star formation
rates (SFRs), dust attenuation and burst mass fractions for 6472 galaxies
observed by the {\it Galaxy Evolution Explorer (GALEX)} and present in
the SDSS DR1 main spectroscopic sample. Parameters are estimated in a
statistical way by comparing each observed broad-band SED (two
\galex and five SDSS bands) with an extensive library of model galaxy
SEDs, which cover a wide range of star formation histories and include
stochastic starbursts. We compare the constraints derived using SDSS
bands {\it only} with those derived using the combination of SDSS and
\galex photometry. We find that the addition of the \galex bands leads
to significant improvement in the estimation of both the dust optical
depth and the star formation rate over timescales of 100~Myr to 1~Gyr
in a galaxy. We are sensitive to SFRs as low as $10^{-3}
M_{\odot}\,{\rm yr}^{-1}$, and we find that low levels of star
formation (SF) are mostly associated with early-type, red
galaxies. The least massive galaxies have ratios of current to
past-averaged SF rates ($b$-parameter) consistent with constant SF over a
Hubble time. For late-type galaxies, this ratio on average decreases
with mass. We find that $b$ correlates tightly with $NUV-r$ color,
implying that the SF history of a galaxy can be constrained on the
basis of the $NUV-r$ color alone.  The fraction of galaxies that have
undergone a significant starburst episode within the last 1~Gyr
steeply declines with mass---from $\sim 20\%$ for galaxies with $\sim
10^8 M_{\odot}$ to $\sim 5\%$ for $\sim 10^{11} M_{\odot}$ galaxies.
\end{abstract}

\keywords{galaxies: evolution-- galaxies: fundamental parameters---
galaxies: starburst---ultraviolet: galaxies}

\section{Introduction}
Modern large-scale galaxy surveys are allowing new constraints to be
placed on the history of star formation (SF) in galaxies.
High-quality optical spectra collected by the {\it Sloan Digital Sky
Survey} (SDSS) have been used to study the recent star formation
histories, dust content and metallicities of over $10^5$ nearby
galaxies (e.g., \citealt{kauff1,kauff2}; \citealt{jarle};
\citealt{tremonti}).  These analyses make use of specific absorption
and emission lines in the galaxy spectra and employ new models of the
spectral evolution of galaxies, which include a physically consistent
treatment of the production of starlight and its transfer through the
interstellar medium (ISM) in galaxies (\citealt{cf}; \citealt{cl};
\citealt{bc2003}).  In addition, the modeling accounts for the
stochastic nature of SF when interpreting galaxy spectra
by using large Monte Carlo libraries of different SF
histories to estimate physical parameters such as stellar mass, age,
and SF rates in a statistical fashion.

In this {\it Letter}, we use a similar approach to interpret the
combined ultraviolet (UV) and optical colors of 6472 SDSS galaxies
observed by the {\it Galaxy Evolution Explorer (GALEX)} \citep{chris}.
We show that the addition of UV information to the optical SEDs of
galaxies leads to significant improvements in the estimates of the
star formation rates (SFRs), starburst histories and dust
attenuations.

\section{Data and Sample}

\subsection{\galex and SDSS data}

We consider galaxies with combined \galex and SDSS photometry, for
which spectroscopic redshifts are available from the SDSS.  \galex
images the sky at far-UV ($FUV$; 1530 \AA) and near-UV ($NUV$; 2310
\AA) bands in two modes: All-sky Imaging Survey (AIS, $m_{\rm lim}(\rm
AB) \approx 20.4$) and Medium-deep Imaging Survey (MIS, $m_{\rm lim}
(\rm AB)\approx 22.7$). Each circular field covers 1.1 sq.\ deg. Here
we use the internal release of the catalog (IR0.2) that consists of
649 AIS and 94 MIS fields, of which 117 and 91 overlap (at least in
part) with the SDSS Data Release One (DR1, \citealt{dr1}). We use
$FUV$ and $NUV$ fluxes and errors derived in elliptical apertures. For
\galex sources with counterparts in the SDSS DR1 spectroscopic sample
we extract SDSS $ugriz$ model colors normalized to the Petrosian $r$
magnitude, 90\% and 50\% Petrosian $i$-band radii, and the likelihoods
of exponential and de~Vaucouleurs $r$-band light profiles.

\begin{deluxetable}{lrrrrr}
\tablecaption{Average Galaxy Parameters and Average Parameter Errors}
\tablehead{ Parameter name & $\langle {\rm G+S} \rangle$ & 
RMS $\Delta({\rm G+S})$\tablenotemark{a} & $\langle \sigma({\rm G+S}) 
\rangle $ & $\langle \sigma({\rm S}) \rangle$ & Gain\tablenotemark{b} }
\startdata
$t_{\rm gal}$            &   6.50  &   0.53  &   1.91  &   2.16  &   $12\%$ \\
$\gamma$                 &   0.40  &   0.06  &   0.21  &   0.22  &    $7\%$ \\
$Z/Z_{\odot}$            &   1.00  &   0.14  &   0.36  &   0.42  &   $12\%$ \\
$\tau_V$                 &   1.24  &   0.17  &   0.60  &   0.86  &   $29\%$ \\
$\mu$                    &   0.36  &   0.03  &   0.14  &   0.15  &    $6\%$ \\
$A_{FUV}$                &   2.03  &   0.23  &   0.62  &   1.05  &   $41\%$ \\
$A_{NUV}$                &   1.46  &   0.18  &   0.46  &   0.78  &   $41\%$ \\
$\log M_*$               &  10.42  &   0.02  &   0.08  &   0.09  &   $11\%$ \\
log SFR(100~Myr)         &   0.00  &   0.07  &   0.27  &   0.54  &   $49\%$ \\
log SFR(1~Gyr)           &   0.18  &   0.12  &   0.30  &   0.55  &   $45\%$ \\
$t_{\rm burst}$          &   2.31  &   0.90  &   1.62  &   1.65  &    $2\%$ \\
$F_{\rm burst}$(100~Myr) &   0.00  &   0.06\tablenotemark{c}  &   0.01  &   0.01  &    $39\%$ \\
$F_{\rm burst}$(1~Gyr)   &   0.02  &   0.08\tablenotemark{c}  &   0.04  &   0.06  &   $39\%$ \\
\enddata
\tablecomments{$t_{\rm gal}$, $t_{\rm burst}$ and $\gamma^{-1}$ are in Gyr, 
SFR in $M_{\odot}\,{\rm yr}^{-1}$, $M_*$ in $M_{\odot}$, and $A_{FUV}$
and $A_{NUV}$ in magnitudes.}
\tablenotetext{a}{RMS change in the parameter value when changing
the frequency of bursts from 50\% to 10\% in the last 2~Gyr in the 
model libraries.}
\tablenotetext{b}{Improvement in $\sigma$ by adding the \galex fluxes
to SDSS $ugriz$ constraints.}
\tablenotetext{c}{Only for galaxies for which $F_{\rm burst}$ is $>0$.}
\end{deluxetable}

\subsection{Sample}

We construct our sample from \galex objects matched to SDSS DR1
objects in a $6''$ radius \citep{mark}. We then restrict the sample to
objects spectroscopically classified as galaxies and with redshifts
$0.005<z<0.25$. To avoid mismatches or matching an unresolved \galex
object with resolved SDSS objects, we additionally impose astrometric
criteria which remove $\approx 10\%$ of the galaxies, leaving 6826.
About 70\% of the remaining galaxies have redshifts in a narrow range
between 0.05 and 0.15 (the number of SDSS galaxies with spectra drops
steeply when $z> 0.1$). \galex MIS and AIS recover on average $80\%$
and $50\%$ of the SDSS main galaxy spectroscopic sample ($r_{\rm lim}<
17.8$). MIS galaxies constitute 60\% of our sample. The addition of UV
selection leads to incompleteness for galaxies redder than $u-r\approx
3.0$ in the MIS and $u-r\approx 2.5$ in the AIS. However, brighter red
galaxies are still present in our sample.

\section{Model Galaxy Parameters}

\subsection{Libraries of galaxy models}

We use the \citet{bc2003} population synthesis code to generate
libraries of stochastic realizations of model SF histories at five
redshifts equally spaced between 0.05 and 0.25.  Each library contains
$\sim 10^5$ models. For simplicity, we follow \citet{kauff1} and
parameterize each SF history in terms of two components: an underlying
continuous model with an exponentially declining SF law
(SFR$[t]\propto\exp[-\gamma t]$) and random bursts superimposed on
this continuous model.\footnote{Specifically, we take the galaxy age
$t_{\rm gal}$ to be distributed uniformly over the interval from zero
to the age of the universe at a given redshift, and the SF timescale
parameter over the interval $0\leq\gamma\leq1\,$Gyr$^{-1}$.  Random
bursts occur with equal probability at all times until $t_{\rm
gal}$. They are parameterized in terms of the ratio between the mass
of stars formed in the burst and the total mass of stars formed by the
continuous model over the time $t_{\rm gal}$. The ratio is taken to be
distributed logarithmically from 0.03 to 4.0. During a burst, stars
form at a constant rate for a time distributed uniformly in the range
30--300~Myr. The burst probability is set so that 50\% of the galaxies
in the library have experienced a burst in the past 2~Gyr.}
Attenuation by dust is described by means of two main parameters (see
\citealt{cf}): the effective $V$-band absorption optical depth
$\tau_V$ affecting stars younger than 10~Myr (that arises from giant
molecular clouds and the diffuse ISM) and the fraction $\mu$ of it
contributed by the diffuse ISM, that also affects older stars.  Based
on previous analyses of SDSS galaxies (\citealt{kauff1},
\citealt{jarle}), we take $\tau_V$ to be distributed from 0 to 6 with
a broad peak around 1 and $\mu$ to be distributed from 0.1 to 1 with a
broad peak around 0.3. Our model galaxies have metallicities uniformly
distributed between 0.1 and 2 $Z_{\odot}$.

\notetoeditor{Please reproduce Figure 1 at 80 percent scale}
\begin{figure}
\epsscale{0.8}
\plotone{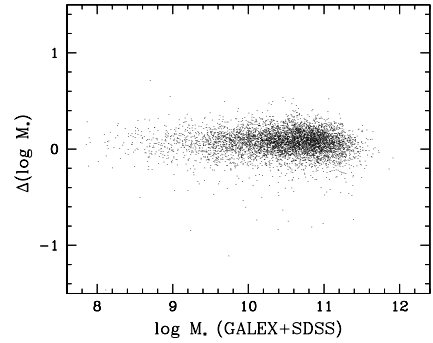}
\caption{ Comparison of galaxy stellar masses derived in \citet{kauff1} 
and in this {\it Letter}. The difference $\Delta(\log M_*)$ is in the sense 
\citet{kauff1} minus this {\it Letter}.}
\label{fig:kauff}
\end{figure}

\notetoeditor{Please place Figure 2 over both columns, and reproduce at 80 percent scale}
\begin{figure*}
\epsscale{0.8}
\plotone{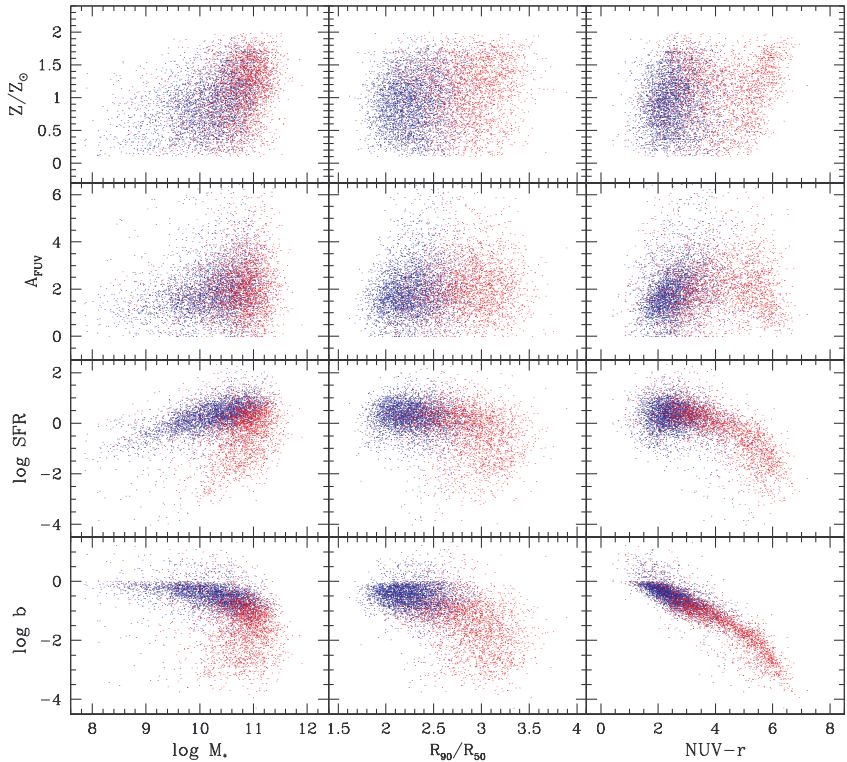}
\caption{ Dependence of metallicity, attenuation and star formation
parameters on fundamental galaxy properties ({\it see
text}).  Blue dots have SDSS profiles that are more likely to be
exponential (late-type galaxies), while red dots have higher
likelihood of $R^{1/4}$ profiles. (Note that this division
may misclassify some galaxies.)}
\label{fig:master}
\end{figure*}

\subsection{Deriving physical parameters}

We derive statistical estimates of various physical parameters by
comparing the observed SED of each galaxy in the sample to the model
SEDs as follows. First, we convert SDSS magnitudes into fluxes,
correcting for the offsets between the SDSS- and AB-magnitude
zeropoints \citep{dr2} and applying small zeropoint offsets to \galex
fluxes. To SDSS flux errors we add calibration errors in
quadrature. After accounting for Galactic reddening, we compare the
SED of the observed galaxy to {\it all} SEDs in the closest-redshift
model library. The $\chi^2$ goodness of fit of each model determines
the {\it weight} ($\propto \exp[-\chi^2/2]$) to be assigned to the
physical parameters of that model when building the probability
distributions of the parameters of the given galaxy.  The probability
density function (PDF) of a given physical parameter is thus obtained
from the distribution of the weights of {\it all} models in the
library. We characterize the PDF using the median and the 2.3--97.7
percentile range (equivalent to $\pm2\sigma$ range for Gaussian
distributions), and also record the $\chi^2$ of the best-fitting
model. We perform this analysis for all galaxies in our sample. The
distribution of $\chi^2$ values of the best-fitting models is
generally very good, implying that our libraries do reproduce the
observed SEDs. However, there is a tail of large $\chi^2$
values--usually identifiable as objects with suspect SDSS magnitudes, or as
``shredded'' objects. We remove $\sim 300$ galaxies with poorest
fits. In the end, we obtain estimates of physical parameters for 6472
galaxies.

\subsection{Parameters and their errors}

We estimate physical parameters in two ways: using the \galex $FUV$
and $NUV$ fluxes combined with the SDSS $ugriz$ fluxes; and using only
the SDSS fluxes. This allows us to quantify the effect of
adding ultraviolet constraints.

Table~1 summarizes our results. For each physical parameter, we list
the sample-averaged parameter value ($\langle {\rm G+S} \rangle$)
derived using {\it GALEX} + SDSS constraints, ``$1\,\sigma$'' estimate
of the average parameter error with (G+S) and without (S) including
the \galex constraints, and the increase in accuracy (gain) achieved
with \galex. We also investigate the effect of changing the fraction
(from 50\% to 10\%) of galaxies in the model libraries that had bursts
over the past 2~Gyr. In Table~1, we report the resulting RMS change in
parameter value. It is usually much smaller than the ``fitting''
error, except for the time $t_{\rm burst}$ since the last burst of SF.

The physical parameters for which the estimates are most significantly
improved with \galex are the 100~Myr- and 1~Gyr-averaged SF rates,
(SFR(100~Myr) and SFR(1~Gyr)), FUV and NUV dust attenuations
($A_{FUV}$ and $A_{NUV}$), and estimates of the fraction of a galaxy's
stellar mass formed in bursts ($F_{\rm burst}$(100~Myr) and $F_{\rm
burst}$(1~Gyr)) over the last 0.1 and 1~Gyr (although it is sensitive
to the assumed frequency of bursts in the model libraries).

\section{Analysis of Physical Properties, Star Formation and 
Starburst Histories}
In Figure \ref{fig:kauff} we compare the stellar masses derived
from our analysis of {\it GALEX} + SDSS colors with those derived
by \citet{kauff1}. These authors used the strengths of the 
${\rm H}\delta_A$ absorption-line index and 4000~\AA\ break in the
SDSS spectra (that trace the recent and past-averaged SF histories)
to improve the constraints on the stellar mass-to-light ratios
relative to estimates based on a single optical color. The scatter in the
difference between the two types of mass estimates (0.11 dex without
$3\sigma$ outliers) is very well matched by the uncertainties in the
two studies (0.08 dex). \citet{kauff1} also compute the fraction of a
galaxy's stellar mass formed in bursts over the last 2~Gyr. In our
study, bursts over 2~Gyr are poorly constrained. This is because the
UV light is less sensitive to stars older than 1~Gyr than the ${\rm
H}\delta_A$ index. We have checked that galaxies for which we find
$F_{\rm burst}{\rm (1~Gyr)}> 0.05$ have 4000~\AA\ breaks and ${\rm
H}\delta_A$ strengths typical of galaxies for which \citet{kauff1}
find $F_{\rm burst}$(2~Gyr)$>0.05$.

We present our main results in Figure \ref{fig:master}, where we show
the dependence of selected physical parameters on three fundamental
galaxy properties: stellar mass, galaxy type, and color. We use the
concentration parameter $R_{90}/R_{50}$ as a proxy for galaxy type.
Values of $R_{90}/R_{50}$ larger than 2.5 correspond mostly to
early-type galaxies, but late-type galaxies can also have large
$R_{90}/R_{50}$ \citep{fuk}. We choose the $NUV-r$ color as it
provides a broad baseline and is not subject to large
k-corrections. Objects are color-coded to indicate whether an
exponential profile (blue) or a de~Vaucouleurs $R^{1/4}$ profile (red)
has higher probability. This serves as an additional rough indicator
of galaxy type.

The first row of panels in Figure \ref{fig:master} presents the
derived stellar metallicities. It was recently demonstrated that
star-forming galaxies in SDSS exhibit a tight metallicity (of ISM)
vs.\ mass correlation \citep{tremonti}. Although metallicity is not
strongly constrained using our method (see Table~1), we do find that
low-mass galaxies do not reach metallicities as high as high-mass
galaxies. (Note that galaxies with $M_*<10^{10} \,M_{\odot}$ are
preferentially disk galaxies, and mostly have $NUV-r<4$.) Massive
galaxies span a wide range in $Z$, with blue and red dots occupying
the same space.

The second row of panels shows that low-mass galaxies on average
suffer less attenuation than massive ones, implying that the dust
content is smaller, but the range of attenuation in more
massive galaxies is larger. We also find that the dust attenuation in
late-type galaxies increases as their $NUV-r$ colors become redder,
but early-type galaxies behave in the opposite fashion. This probably
indicates that the reddest early-type galaxies have little gas,
ongoing SF and dust.

Next we examine the total current SFR (averaged over the last 100
Myr). We find that \galex is sensitive to very low SF levels, 
$\sim10^{-3} M_{\odot}\,{\rm yr}^{-1}$. Not surprisingly, galaxies
with the lowest SFRs are primarily of early type. However, there is
an interesting concentration of red dots at $\log\,{\rm SFR}\sim
0.5$.  This ``population'' has lower concentration index and bluer
$NUV-r$ color than low-SF early-type galaxies. Inspection of the 
images of a subset of these galaxies with high metallicities reveals
that about half of them are spirals (often disrupted), while most 
others show (low surface-brightness) disks. Some 10\% appear to be
true ellipticals. This emphasizes the crudeness of classification 
based on profile fitting. Finally, we note that we compared our 
SFRs with those derived from aperture-corrected nebular emission
\citep{jarle} and found excellent overall agreement. A more detailed
comparison of SF metrics will be presented elsewhere.

The last row shows the ratio of current (last 100~Myr) to
past-averaged SF (the ``$b$-parameter''). Galaxies with constant SFR
will have $\log\,b=0$, while those undergoing a burst can have
$\log\,b>0$. We find that $b$ is close to unity for low-mass galaxies
but extends to 0.1 for massive, late-types galaxies, in agreement with
the results of \citet{jarle}. Early-types reach values as low as
$b\approx 3\times 10^{-4}$. The tight correlation between $b$ and
$NUV-r$ color is most remarkable. It implies that the SF history of a
galaxy can be constrained on the basis of the $NUV-r$ color alone.

\notetoeditor{Please reproduce Figure 3 at 80 percent scale}
\begin{figure}
\epsscale{0.8}
\plotone{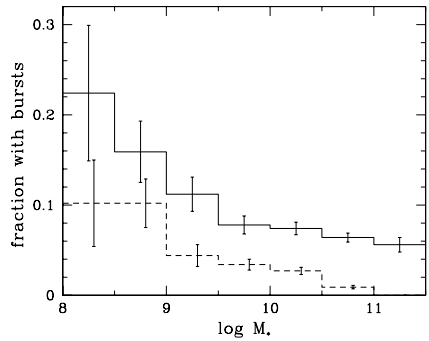}
\caption{ Fraction of galaxies of a given mass with starbursts.
Histograms correspond to fraction of galaxies with bursts (that
produced more than 5\% of stellar mass) having occurred in the last
100~Myr and 1~Gyr (dashed and filled lines) and . The fraction of
galaxies with recent bursts declines sharply with galaxy mass. }
\label{fig:fburst}
\end{figure}

We have also examined the relation between SFR and FUV attenuation
for the galaxies in our sample. We find that increasing SFRs are always
associated with larger attenuations. At a fixed attenuation, however,
early-type galaxies can have SFRs several orders of magnitude
smaller than late-type galaxies. This suggests a different
dependence between dust content and SF in the two types of galaxies.
\citet{buat} also find that the dust attenuation increases with
the amount of recent SF as estimated by the dust-corrected NUV flux.

One might worry that because our sample has UV selection, the results
shown in Figure \ref{fig:master} may not apply to the galaxy
population as a whole. We have analyzed the relations between the same
galaxy properties as in Figure \ref{fig:master} for the MIS fields,
where the UV coverage of galaxies in the SDSS spectroscopic sample is
substantially more complete. We found exactly the same {\em relations}
between SFR, $b$, $A_{FUV}$ and mass, concentration and color. The
most notable difference is that the MIS data (by virtue of going
deeper) contain a larger fraction of massive, concentrated red
galaxies. Therefore, while the above results are useful for
identifying various trends, they cannot be used to study the relative
proportions of galaxies with different properties.

Finally we perform a time-resolved analysis of starbursts. We select
galaxies that have experienced starburst in which more than 5\% of
stellar mass ($F_{\rm burst} \geqslant 0.05$) was formed over the last
100~Myr and 1~Gyr. Figure \ref{fig:fburst} gives the fraction of
galaxies {\it of a given mass} that have undergone a recent
starburst. The dashed histogram corresponds to bursts within the last
100~Myr. For $\log M_*< 9$, the fraction is $\sim 10\%$, but it
declines to zero for the most massive galaxies. Over the last 1~Gyr
(solid histogram) some 20\% of low-mass galaxies having experienced
bursts. This fraction falls to 5\% for larger masses. Qualitatively
consistent results are obtained if we restrict the analysis to the MIS
fields. Similar results were obtained by \citet{kauff2}, who analyzed
starbursts over a 2~Gyr timescale.

In this preliminary study of SF history using the UV photometry from
\galex, we demonstrate the promise and point to possible limitations
of this dataset when interpreted with galaxy models. As we gather more
data we should be able to better explore the trends suggested here,
and to get better statistics on low-mass galaxies.

\acknowledgments 
We thank Sukyoung Yi for comments. SS thanks Andrew Gould. 
SC thanks the Alexander von Humboldt Foundation.
\galex is a NASA Small Explorer, launched in April 2003.  We
gratefully acknowledge NASA's support for construction, operation, and
science analysis for the \galex mission, developed in cooperation with
the CNES of France and the Korean Ministry of Science and Technology.
Funding for the creation and distribution of the SDSS Archive has been
provided by the Alfred P. Sloan Foundation, the Participating
Institutions, the NASA, the NSF, DoE, Monbukagakusho, and the Max
Planck Society.


\end{document}